\documentclass[11pt]{article}
\usepackage{amssymb,cite,epsf,graphics,graphicx}
\usepackage{amsmath}
\usepackage{amsthm}
\usepackage{amscd}
\newcommand\blank[1]{}

\newcommand{\fract}[2]{{\textstyle\frac{#1}{#2}}}

\newcommand\eq{\begin{equation}}
\newcommand\en{\end{equation}}
\newcommand\bea{\begin{eqnarray}}
\newcommand\eea{\end{eqnarray}}
\newcommand\nn{\nonumber}
\newcommand\ba{\(\begin{array}}
\newcommand\ea{\end{array}\)}

\newcommand{\resection}[1]{\setcounter{equation}{0}\section{#1}}

\newcommand{\One}{{\hbox{{\rm 1{\hbox to 1.5pt{\hss\rm1}}}}}}

\begin{document}
\begin{titlepage}
\vskip 0.5cm
\begin{flushright}
DCPT-03/37 \\
{\tt hep-th/0308053} \\

\end{flushright}
\vskip 1.5cm
\begin{center}
{\Large{\bf Finite lattice Bethe ansatz systems and the Heun
equation }}
\end{center}
\vskip 0.8cm
\centerline{Patrick Dorey$^1$,
Junji Suzuki$^2$
and Roberto Tateo$^{1,3}$}
\vskip 0.9cm
\centerline{${}^{1}$\sl\small Dept.~of Mathematical Sciences,
University of Durham, Durham DH1 3LE, UK\,}
\vskip 0.2cm
\centerline{${}^{2}$\sl\small Dept.~of Physics, Faculty of Science,
Shizuoka University, Ohya 836 Shizuoka, JP\,}
\vskip 0.2cm
\centerline{${}^3$\sl\small Dipartimento di Fisica Teorica,
Universit\`a di Torino, Via P. Giuria 1, 10125 Torino, I\,}
\vskip 0.3cm
\centerline{e-mails: {p.e.dorey@dur.ac.uk},
{sjsuzuk@ipc.shizuoka.ac.jp},
{roberto.tateo@dur.ac.uk}}

\vskip 1.25cm
\begin{abstract}
\noindent
We study the   P{\"o}schl-Teller  equation in complex domain and
deduce infinite families of  $TQ$ and Bethe ansatz equations,
classified by four  integers. In all these models the form of $T$
is very simple, while  $Q$ can be explicitly written in terms of
the Heun  function. At particular values there is a interesting
interpretation in terms of finite lattice spin-$\frac{L-2}{2}$ XXZ
quantum chain with  $\Delta= \cos \frac{\pi}{L}$ (for
free-free boundary conditions), or
$\Delta=-\cos\frac{\pi}{L}$ (for periodic boundary conditions).
 This result  generalises  the
findings of Fridkin, Stroganov and Zagier. We also  discuss the
continuous (field theory) limit of these systems  in view of the
so-called ODE/IM correspondence.
\end{abstract}
\end{titlepage}
\setcounter{footnote}{0}
\def\thefootnote{\fnsymbol{footnote}}
%
\resection{Introduction}
\label{intr}
Some  years ago, an unexpected connection was found
between  certain $0+1$ dimensional quantum-mechanical problems
 and $1+1$ dimensional conformal
field theories~\cite{DT1,BLZ,JS1,Fe,DT2,JS2,JS3,DDT1}.
The simplest
example
involves  the
Schr{\"o}dinger equation
\eq
-\frac{d^2}{dx^2} \psi(x,E) + (x^{\alpha}-E)  \psi(x,E)=0~,
\label{sch}
\en
and the fact that (\ref{sch})
has  unique solution  $y(x,E)$, entire in $x$ and $E$,
which  decays along the positive
real axis as   $x \rightarrow \infty$.
The  function  $y(x,E)$ can be shown \cite{DT2}
to satisfy a Stokes  relation
\eq
T(E) y(x,E)= \omega^{-1/2}\ y(\omega x, \omega^{-2} E ) +
\omega^{1/2}\ y(\omega^{-1} x,
\omega^2 E)~,
\en
with  $\omega=\exp(i 2\pi/(\alpha+2))$. This is strikingly similar
to the  $TQ$ relation, a functional equation which was introduced
in the context of the six-vertex model by Baxter (cf.
\cite{Bbook}).

The correspondence is actually much more precise:
the Stokes multiplier~\cite{Sh} $T(E)$ and the spectral determinants~\cite{V1}
 $Q^+(E)=y(x,E)|_{x=0}$ and $Q^-(E)=y'(x,E)|_{x=0}$ are
equal to the     ground-state eigenvalues of the  transfer matrix
and of the
Q operators, respectively, of the six-vertex  or, equivalently, of the
spin-$\fract{1}{2}$ XXZ quantum chain, in suitable continuum
(field theory) limits. The continuum limit of the six-vertex model
is related to conformal theory at $c=1$, and the description of
such theories in a framework similar to Baxter's lattice setup was
addressed in an important series of works by Bazhanov, Lukyanov
and Zamolodchikov~\cite{BLZ1,BLZ2,BLZ3}. The correspondence
also applies, directly, in this context. Since the initial
observation of \cite{DT1}, many mathematical aspects of the
correspondence have been clarified~\cite{BLZ,JS1,DT2}, but the
physical reasons for its existence remain  mysterious. In
addition, up to now, the correspondence has been limited to the
ground-state energy of the conformal field theory and it is
unclear whether it can be extended to massive theories, to excited
states\footnote{While we were finishing writing this article the
preprint~\cite{BLZodeex} appeared. In this very interesting paper,
Schr{\"o}dinger equations corresponding to excited
states are proposed.}, or  to  finite lattice
systems\footnote{An extension of the finite lattice Baxter $TQ$
relation with extra coordinates like parameters has been recently
introduced by Weston and Rossi in ~\cite{RW} for $q$ generic (see also 
\cite{CK1,CK2} for $q$ a root of unity). The relationship
between these results  and the ODE/IM
 correspondence still needs to be clarified.}.

The purpose of this paper is to report some progress on the last
of these questions. The key difference in  finite lattice problems
is the appearance of an extra function $\Phi(E)$  in the $TQ$
relation:
\eq
T(E)Q(E)= b^{-1} \ \Phi(\omega E) Q(\omega^{-2} E)+  b \
\Phi(\omega^{-1} E) Q(\omega^2 E)~.
\en
($b$ is a pure phase.) This function is fixed by the problem under
consideration. In particular, it encodes the number of lattice sites,
and tends to $1$ when a suitable continuum limit is taken.

 We start with the fact that at the points $\alpha=2$ and
$\alpha=1$  the solutions of (\ref{sch}) can be given explicitly,
in terms of hypergeometric functions. In the search for
finite-lattice generalisations of the correspondence it seems
natural to begin with these two models. At  the linear point,
$\alpha=1$, there is the additional  advantage that the r{\^o}les
of $E$ and $x$ are interchangeable: setting $z=x-E$, (\ref{sch})
becomes the standard Airy equation
\eq
-\frac{d^2}{dz^2} \psi(z) + z  \psi(z)=0~,
\label{airy}
\en
and therefore~\cite{Fe,DT1}
\eq
Q^+(E)|_{\alpha=1}=\hbox{Ai}(-E)~~\hbox{and}~~~ Q^-(E)|_{\alpha=1}=
-\hbox{Ai}'(-E)~,
\en
which means that   the functions  $Q^{\pm}(E)$ themselves satisfy
differential  equations.

The ideal situation would be that the
finite-lattice version of the problem would share
this  $(E,x)$-democracy.
Surprisingly, this was precisely the discovery made by
Fridkin, Stroganov and Zagier in\cite{FSZ}:
though they did not make a connection with the
earlier results of~\cite{DT1,BLZ,JS1,Fe,DT2},
they empirically discovered  that the  $Q$ function for the spin-$\frac{1}{2}$
XXZ quantum  chain with free-free boundary conditions and $\Delta=1/2$
was related to a specialisation of the  P{\"o}schl-Teller equation
\eq
\left (-
 \frac{d^2}{d s^2} - \frac{9 n(2n+1)}{ 2\cosh^2 3s/2} \right )
\chi(s)
 =- \chi(s)~.
\label{FSZ}
\en
In a subsequent paper\cite{St1}, the same spin chain, but
with $\Delta=-1/2$ and with periodic boundary conditions,
was also related to the  P{\"o}schl-Teller equation
with a  different eigenvalue.

In this paper we show that the most general P{\"o}schl-Teller
equation (given in equation (\ref{PT-schro}) below) contains
infinite families of finite $TQ$ and Bethe ansatz equations,
selected by fixing the four parameters $(M,N,L,m)$ to positive
integer values. At the  particular values $(0,N,L,0)$ and
$(0,N,L,1)$ there is a straightforward and interesting
interpretation in terms of finite lattice spin-$\frac{L-2}{2}$ XXZ
quantum  chains.

The plan of the paper is as follows.  The
relevant equation and its analytic properties   are discussed in
section~\ref{Strog}. The $TQ$ relation is derived in
section~\ref{Stokes} and the relationship with the quantum spin
chains discussed  in section~\ref{model}. In section~\ref{baesect}
some numerical results are reported and in  section~\ref{cl}  the
continuous limit of the equation is studied in view of the ODE/IM
correspondence. Finally section~\ref{con} contains our
conclusions. There are two appendices: in appendix~\ref{Heun} the
solution in terms of the Heun function is derived, while  the
locations of the trivial zeroes of the  solution are discussed in
appendix~\ref{fixtrivialzeros}.

\resection{The differential equation}
\label{Strog}
We consider the generalised  P{\"o}schl-Teller  equation
\begin{equation}
\left (
   -\frac{d^2}{d s^2} - \frac{N(N+1)}{ \cosh^2 s}
+  \frac{M(M+1)}{ \sinh^2 s}    \right  ) \chi(s)
 =
 -\sigma^2 \chi(s ).
\label{PT-schro}
\end{equation}
As explained in appendix \ref{Heun},  the differential equation
(\ref{PT-schro}) can be mapped into the Heun equation, allowing
any  solution of (\ref{PT-schro}) to be written in terms of the
Heun  function~$H$.
This is one of the reasons why the P{\"o}schl-Teller equation has
historically played an important r{\^o}le in the
quantum-mechanical modelling of two-body problems with short-range
interactions. In these applications the
wavefunction is usually defined on the real axis, and
the physical
requirement of square integrability constrains $\sigma$ to integer
values, allowing the  solution to be written in  terms of
 the more standard ${}_2 F_1$ hypergeometric function.

In this paper we shall instead consider equation (\ref{PT-schro})
on the whole complex plane, and one of the requirements placed on
its solutions will be meromorphicity. The demand that $\chi(s)$ be
single-valued around the singularities of $\cosh^{-2}s$ and
$\sinh^{-2}s$ immediately restricts the parameters $N$ and $M$ to
integer values. However, in the following   we shall  impose
further conditions, which emerge as follows.

Introduce a new variable  $x^L =-\exp(2s)$\,, and set
\eq
m=\sigma L-1~,~~~~\psi_{M,N,m}(x)= x^{(m+1)/2} \chi(\ln(\sqrt{-x^L}))~.
\en
Then $\psi_{M,N,m}(x)$ is solution of
\begin{equation}
\left( x^2 \frac{d^2}{dx^2} -m x
\frac{d}{ d x} -\frac{L^2  N(N+1) x^L}{(x^L-1)^2}
  +\frac{L^2  M(M+1) x^L}{(x^L+1)^2}
   \right ) \psi_{M,N,m}(x) =   0~.
 \label{odegauge}
\end{equation}
The requirement that $\psi_{M,N,m}(x)$ be single-valued on the
whole  complex plane  leads to the quantisation of the four
parameters  $N,M,L,m$ to integer values. To see the quantisation
of $m$, notice that the points $x=0$ and $x=\infty$ are, in
general, singular points of  (\ref{odegauge}), and in their
vicinity solutions behave as
\eq
 \psi_{M,N,m}(x) \sim \alpha+ \beta x^{m+1} + \dots~~,
\label{asy}
\en
and therefore single-valuedness  constrains $m \in \mathbb{Z}$.
The case $m \ge 0$ is already very rich in structure, and so
we shall restrict ourselves to this case.
Without any further loss of generality, we   conventionally set
 $L, N,M \ge 0$.
All of
this was to allow the
single-valuedness and hence meromorphicity of
the solutions to (\ref{odegauge}).
To single out
one particular solution, to play the r\^ole of the
function $y$ in (\ref{sch}),
we shall impose the boundary
condition
\eq
\psi_{M,N,m}(x)|_{x \sim 1} \sim (1-x)^{N+1}~.
\label{zp}
\en
This condition is natural, in that (\ref{odegauge}) has
regular singularities  at
\eq
(x^L \mp 1)|_{x=x_{i,\pm}}=0~.
\en
Near these points a generic
solution behaves  as
\bea
\psi_{M,N.m}(x) &\sim&  c (x-x_{i-})^{-N}~,~~~(x^L \sim 1)~, \\
\psi_{M,N,m}(x) &\sim&  \tilde{c}  (x-x_{i+})^{-M}~,~~~(x^L \sim -1)~.
\label{poles}
\eea
If we impose $c=0$ then, exceptionally,
\eq
\psi_{M,N,m}(x) \sim  d ( x-x_{i-})^{N+1}~,
\label{con1}
\en
which up to the (arbitrary) normalisation
is exactly the  condition (\ref{zp}).
The choice to impose boundary conditions near regular singularities in this way
 might
seem
to be unmotivated at this
stage, but it will be crucial in making a connection
with (\ref{sch}) in
the scaling limit. We shall return to this point in section~\ref{cl}.

 Notice that  equation  (\ref{odegauge}) is   invariant under the
transformation
\eq
(x, M, N) \rightarrow  ( x \omega^{1/2}, N, M)~,
\label{rotate2}
\en
where $\omega = \exp(2i \pi/L)$, and consequently also under
\eq
 (x,M,N) \rightarrow  (x \omega , M, N)~.
\label{rotate1}
\en
These symmetries force further
zeroes in  $\psi_{M,N,m}(x)$.
Being images of $x=1$ under certain rotations, they are
located on the unit circle, and are, in some sense, trivial.
They will, however,
contribute non-trivially to the Bethe ansatz equations which
fix the nontrivial zeroes -- see, for example, (\ref{defq}) below.
The determination of the locations of the trivial zeroes
is simple but technical, and
we relegate it to appendix~\ref{fixtrivialzeros}.
\resection{The connection formula and the $TQ$ relation}
\label{Stokes}
We shall now formulate the problem in a setup  similar to that
used in presence of  Stokes sectors~\cite{Sh}.
Set  $g(x)= \psi_{M,N,m}(x)$ and define
\eq
 g_{k}(x)=g({\omega}^k x)~.
\en
Then the  symmetry   (\ref{rotate1}) ensures that
$g_1(x)$ and $g_{-1}(x)$ are also solutions of equation   (\ref{odegauge}).
Near $x=1$ they behave as
\eq
g_1(x) \sim c_{+} (x-1)^{-N}~~,~~g_{-1}(x) \sim c_{-} (x-1)^{-N}~,
\en
and so the pair of  functions $\{ g_0(x),g_{1}(x) \}$ is (apart from the
particular values of $m=L-1~ {\rm mod}\, L$) a basis of solutions.
Expanding $g_{-1}$ in this basis and rearranging,
\eq
W[-1,1] g_0(x) =W[-1,0] g_1(x) + W[0,1] g_{-1}(x)~,
\label{tq1}
\en
where the Wronskian $W[i,j ]$ is
\eq
W[i,j ] = g_i(x) {g'}_j (x)-{g'}_i(x)g_j(x)~.
\en
Because of the term  $-m x \frac{d g(x)}{dx}$ in (\ref{odegauge})
one can deduce that the Wronskian between  any pair of  solutions has the form
\eq
W[g,f] =cst~x^{m}~,
\en
and one  can factorise $x^m$ out of  (\ref{tq1}).
We can now use the large $x$ asymptotic  (\ref{asy})
\begin{equation}
g (x) \sim a + bx^{m+1} +\dots
\label{asym}
\end{equation}
to find the exact expression for $W[-1,0]$ and $W[0,1]$. For $a,b \ne 0$
the result is
\bea
W[-1,0] &=& 2i ab(m+1) {\omega}^{(m+1)/2} \sin(\fract{m+1}{L}\pi) x^{m}~~~~, \\
W[0,1]  &=& 2i ab(m+1) {\omega}^{-(m+1)/2}   \sin(\fract{m+1}{L} \pi) x^{m}~~,
\\
W[-1,1] &=&4i ab(m+1) \sin(\fract{m+1}{L}\pi)\cos(\fract{m+1}{L} \pi) x^{m}~,
\eea
and (\ref{tq1}) becomes
\eq
2 \cos(\fract{m+1}{L} \pi) g(x)=
{\omega}^{\fract{m+1}{2}} g({\omega}^{-1} x )+
{\omega}^{-\fract{m+1}{2}} g({\omega} x)~.
\label{stokes}
\en
This equation is almost
identical to Baxter's  $TQ$ relation,  save for the fact
that  $Q(x)$  is, by definition, entire 
while  $g(x)=\psi_{M,N,m}(x)$ is not.
This can be simply overcome by 
introducing a new   function $q(x)$ defined as
\eq
(x^L-1)^N  (x^L+1)^M  g(x) =\prod_{j=0}^{\ell-1} (x-(\omega')^j)^{2N+1}
\prod_{i=1}^{N_k} (x-(\omega)^{\frac{2k_i+1}{2}})^{2M+1} q(x)~.
\label{defq}
\en
In the above we have used the knowledge of the trivial zeroes and poles
discussed in appendix~\ref{fixtrivialzeros}. Note that $\ell$,
$\omega'$,  $k_i$ and $N_k$ are respectively defined in
(\ref{ell}), (\ref{omega1}), (\ref{ki}) and at the very end of
appendix~\ref{fixtrivialzeros}.

By a consideration of the possible singularities and trivial
zeroes in the previous section and in
appendix~\ref{fixtrivialzeros}, we immediately deduce the
following factorised form for $q(x)$:
\eq
q(x)= \prod_{i=1}^K \left( 1- \frac{x}{x_j} \right),
\en
with
\eq
x_{j} = 1/x_{K+1-j}~.
\en
The function $q(x)$ also  satisfies a  
$TQ$-type  relation, and  it is entire.

The number of nontrivial zeroes, $K$, of $q(x)$  is easily evaluated
 by noticing that  $\psi_{M,N,m}(x)$ is a meromorphic function of $x$.
Then, the  asymptotic behaviour (\ref{asym})  indicates that
${\rm Max}(m+1,0)$ should be equal   to the total number of zeroes minus
the total number of poles existing at finite $x$. For   $m+1>0$,
this leads to
\eq
K= (m+1)+N(L-\ell)+M(L-N_k)-\ell(N+1)-N_k(M+1)~.
\label{Kappa}
\en

The set of numbers $\{x_j\}$ constitutes the
nontrivial zeroes of the
wavefunction. They
are fixed by the Bethe ansatz equations.
To match the standard notation, we change variables
$x \rightarrow -x$ and $E_j=-x_j$, and also set
$x=\exp(u)$, $E_j=\exp(u_j)$.
In these new variables
the connection ($TQ$) formula (\ref{stokes}) becomes
\eq
\tau(u) Q(u) ~=~
{\phi}(u-2i\eta ) Q(u-2i\eta)+ {\phi}(u+2i\eta) Q(u+ 2i\eta)~,
\label{tqf}
\en
where we defined
\bea
Q(u) &=& \prod_{j=1}^K \sinh(\frac{u-u_j}{2})~,
\label{QQ} \\
{\phi}(u)&=& \prod_{j=0}^{\ell-1} \cosh^{2N+1}(\frac{u}{2}-i\eta' j)
      \prod_{i=1}^{N_k}   \cosh^{2M+1}(\frac{u}{2}-i\frac{2 k_i+1}{2}\eta )~,
\label{PhPh}
\\
\tau(u)&=& (-1)^{N+M} 2 \cos(\fract{m+1}{L} \pi)  \phi(u)~,
\label{taut}
\eea
with $\eta=\pi/L$ and $\eta'=\pi/\ell$.

From its explicit form, $\tau(u)$ should be pole-free while the formal solution
of the above algebraic equation seems to possess poles at the
zeroes of $Q(u)$.
Thus the residue at these points must be vanishing.
This is exactly the same reasoning which leads to the Bethe ansatz equation
in integrable systems.
A suitably-chosen  solution to the resulting Bethe ansatz equation 
characterises $Q(u)$, and consequently $q(x)$, 
and
exhibits several interesting patterns depending
on the choice of parameters.
Before presenting  examples, however, we shall discuss the
connection of our findings to  quantum magnets.

\resection{Model identification}
\label{model}
Consider a one-dimensional quantum system
in which quantum spins
of magnitude $S$ are
assigned to each site of a length ${\cal N}_S$ chain.
They interact via spins of magnitude 1/2 living on bonds.
The strength of the interaction is characterised by
$\triangle = \cos \lambda$.
Assume further either periodic boundary conditions
(p.b.c), or free-free (f-f) boundary conditions
with  $U_q(sl_2)$ invariant interaction and $q=\exp i\lambda$.
For example, the
 Hamiltonian for $S=\frac{1}{2}$\,, and periodic boundary conditions, is:
$$
{\cal H} = \sum_{n=1}^{{\cal N}_{1/2}} 
 \bigl (
          \sigma^+_n \sigma^-_{n+1} +\sigma^-_n \sigma^+_{n+1} +
          \frac{\Delta}{2} \sigma^z_n \sigma^z_{n+1}
    \bigr )\,.
$$

The transfer matrix $T_S(u)$ is given by either a single
(p.b.c)  or
doubled (f-f)  form via Skylanin's construction\cite{Sklyanin}.
The auxiliary space has spin $\frac{1}{2}$\,, and
the quantum space is given by the ${\cal N}_S$-fold tensor product
of a spin $S$
space.
Then the following
$TQ$ relation holds:
\begin{eqnarray}
T^{(r)}_S(u) Q_S(u) &=& \phi_S(u-2i  S \lambda )Q_S(u+2 i\lambda)+
   \phi_S(u+2 i  S\lambda )Q_S(u-2 i\lambda)~,  \label{spinTQ1}\\
  Q_S(u) &=&
\begin{cases}
      \prod_j  \sinh \frac{u-v_j}{2}~ , & {\rm  (p.b.c)}   \\
      \prod_j  \sinh \frac{u-v_j}{2}\sinh \frac{u+v_j}{2}~,
   & {\rm  (f-f)}   \\
 \end{cases}
    \nonumber \\
  \phi_S(u) &:=&
\begin{cases}
 \prod_{\alpha=1}^{{\cal N}_S}  \sinh\frac{(u-\omega_{\alpha})}{2}~, 
& {\rm  (p.b.c)} \\
 \sinh(u) \prod_{\alpha=1}^{{\cal N}_S}
   \sinh\frac{(u-\omega_{\alpha})}{2} \sinh\frac{(u+\omega_{\alpha}) }{2}~,
 &{\rm (f-f)}
\end{cases}
\nonumber
\end{eqnarray}
where  $\omega_{\alpha}$ stands for the inhomogeneity,
and
$T^{(r)}_S(u)$ stands for
the renormalised  transfer matrix:
$$
T^{(r)}_S(u)=
\begin{cases}
      T_S(u)~ & {\rm  (p.b.c)}   \\
      \sinh u~ T_S(u)~
   & {\rm  (f{-}f)}   \\
\end{cases}
$$

For periodic boundary conditions, 
the above relation can be shown
directly, while for free-free
boundaries, it generalises established results
for $S=\frac{1}{2}$ and~$1$ \cite{MezNep, YB}.

The similarity between (\ref{spinTQ1}) and
 our connection formula
(\ref{tqf}--\ref{taut}) is clear.
To check the precise  correspondence,
we now examine some
simple examples, taking
$M=0$ and $m=0$ or $1$. For $m=1$, we additionally
impose  that $L$ be odd, so that in all cases $\ell=1$.
Then  ${\phi}$  in (\ref{PhPh}) simplifies considerably:
$$
{\phi}(u)=
\begin{cases}
\cosh^{2 N+1}(\frac{u}{2})~, & (N_k=0) \\
\frac{1}{2} \sinh (u) \cosh^{2 N}(\frac{u}{2})~, & (N_k=1)\,.
\end{cases}
$$
Noting also the property
\eq
\cosh(\frac{u}{2}  \pm i \fract{\pi}{L}) =
\pm i \sinh(\frac{u}{2} \mp  i \fract{L-2}{2L}\pi)
= \pm i
\sinh(\frac{u}{2} \mp i (L-2) \fract{\eta}{2})~,
\en
for $N_k=1$
it is
immediately seen that the connection rule 
(\ref{tqf}--\ref{taut}) 
coincides with
(\ref{spinTQ1}) for the spin $\frac{L-2}{2}$ chain with f-f
boundaries, an even number of sites ${\cal N}_S = 2N$,
and with parameters
 $\eta=\lambda=\frac{\pi}{L}$, $u_j=v_j$, and
$\omega_{\alpha}=0$. For the match to be complete the  function $\tau(u)$ 
should be related to an  eigenvalue  $T^{(r)}_S(u)$
of the   transfer matrix as   $ 2 (-1)^{N+1} \tau(u)= T^{(r)}_S(u) $.

Similarly, for $N_k=0$ and $L$ odd, the connection rule
coincides with 
(\ref{spinTQ1}) for the spin $\frac{L-2}{2}$ chain, but  with
p.b.c and an odd number of
sites ${\cal N}_S = 2N+1$. The
parameters must be identified as $\lambda=\pi-
\eta=\frac{(L-1)\pi}{L}$, $ u_j=v_j+\pi i$,
$(-1)^{\frac{L}{2}+m-1}  \tau(u)=T^{(r)}_S(u+\pi i)$,
and $\omega_{\alpha}=0$.

Specialising to $L=3$, the above results recover the
findings of \cite{FSZ, St1}.
The coincidence between the ODE and the spin
chain was checked numerically
for $N_k=0$, $L=3, 5$,
with $N=1,2,3$.
We adopted a `brute force' diagonalisation of the transfer
matrices associated to the spin systems, and then verified that
the resultant spectra contain eigenvalues of the form 
 $T^{(r)}_S(u) = 2 (-1)^{N+1} \tau(u)$.
These correspond to the particular solutions of the Bethe ansatz equations
 (\ref{bae}) which will be
associated with the ODE in  the next section.
The eigenvalues are not particular members of the spectra:
they are neither  the largest in magnitude nor the smallest.
However, the same BAE patterns do play a distinguished  r{\^o}le 
in a particular (isotropic)   fused model. 
See the discussion in section \ref{con}.
The remarkable simplicity of the expression for $T_S(u)$
comes from the elementary expression for $\tau(u)$ in (\ref{taut});
it reflects the special nature of the points we are examining 
even before the scaling limit is taken.

When $N_k=0$ and $L$ is even, our connection rule differs from
the periodic boundary condition case of
(\ref{spinTQ1}) by a sign. This
suggests the need for a different choice of boundary
conditions for the spin model.
We leave this for future work, as well as the identification
of the connection rule  and the $TQ$ relation in higher spin chains,
with general choice of $N,M, L, m$,
where the inhomogeneities $\omega_{\alpha}$ should be chosen properly.

We make one further, more general,
remark in closing
this section.
There are
some ambiguities in the choice of functions
in the lattice model and the ODE\,: in particular,
equation~(\ref{stokes}) is invariant if
$\tau(u)$ and $\phi(u)$ are multiplied 
by a common arbitrary entire function of $x^L$
\footnote{
The explicit representation-theoretical construction of the $Q$ operator
is not our concern here. However, we  mention  the recent work \cite{CK1, CK2},
 where the subtleties which arise for $q$ a root of unity are addressed, 
taking into account   so-called ``exact complete strings" \cite{FK}.
It is worth noting that these strings are related to the above-mentioned  
possibility to
multiply (\ref{stokes}) by an entire function of $x^L$.
 }.
Although the choice we have made above appears to be the most 
natural, and is supported by our numerical results, we 
cannot exclude the possibility
of the extra factor being relevant in a more general situation.
We hope to be able to
resolve this issue in a future publication.
\resection{The Bethe ansatz equations and string  like solutions}
\label{baesect}
{}From equation (\ref{tqf}), and the reasoning given after that equation,
the zeroes $\{ u_j \}$ of $Q(u)$
satisfy the following  Bethe Ansatz equations (BAE)\footnote{
Takemura\cite{Takemura} has also discussed the application of
BAE to
the determination of zeroes of wave functions
for the generalised P{\"o}schl-Teller equation
(or the $BC_1$ Calogero-Sutherland model), however,  with the $L_2$ property.
The BAE itself is  similar  to the semi-classical form,
 thus different from that described here, the quantum form.}
\begin{equation}
\frac{{\phi}(u_j-2i\eta ) Q(u_j-2i\eta)}
     {{\phi}(u_j+2i\eta)  Q(u_j+2i\eta)} =-1~,~~(j=1, \cdots, K)~.
\label{bae}
\end{equation}
The solutions to these equations which are related to our ODE 
exhibit various interesting patterns of zeroes
depending on the choice of $(M,N,L, m)$, and in this section
we comment on a few specific examples.
First we take one of $N, M$ to be zero.
We start with the
$M=0$ case, which is a natural extension of that treated
in~\cite{FSZ}.

\begin{table}[htbp]
\begin{center}
\begin{tabular}{|c|c|}
\hline
$\pm$ 1.17053198009 $\pm$ 0.82128068487  i &
      $\pm$ 0.74741179593 $\pm$ 0.80157843231  i \\
$\pm$ 0.51014119671 $\pm$ 0.79624585390  i &
      $\pm$ 0.33682705394 $\pm$ 0.79395490465  i \\
$\pm$ 0.19297842512  $\pm$ 0.79284699715   i&
     $\pm$ 0.06295373414 $\pm$  0.79238018877  i  \\
\hline
\end{tabular}
\caption{BAE roots for $L=4$, $m=0$, $N=12$.}
\label{tableroots}
\end{center}
\end{table}

When $m=0$ or $1$,
the  BAE roots assume the famous string patterns of length 
$L{-}2$; the number of strings is generically $N$.
We confirmed that  this leads  to a proper solution  of the differential
equation  (\ref{PT-schro}). This means that $\psi_{M,N,m}(x)$
is, modulo a trivial change of variable,  directly related to the
ground-state expectation value of the operator $Q$.
As an example,
the set of Bethe ansatz roots 
for  $L=4$, $m=0$, and $N=12$
is given in table~\ref{tableroots}.

\begin{figure}[hbtp]
\centering
\includegraphics[width=6cm]{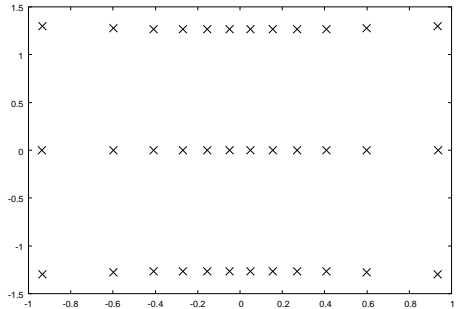}
\caption{ \small
Zeroes of $Q$   (\ref{PT-schro}) at $(M,N,L,m)=(0,12,5,0)$,
which illustrates the 3 string solution.
}
\label {3string}
\end{figure}

\begin{figure}[hbtp]
\centering
\includegraphics[width=6cm]{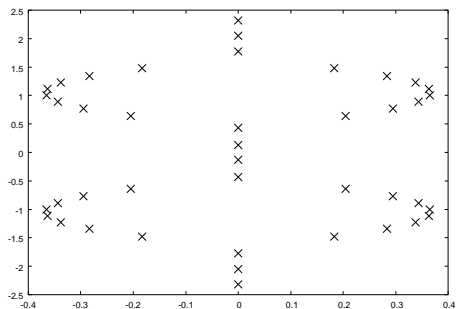}
\caption{ \small
Zeroes of the wavefunction of   (\ref{PT-schro}) with
$(M,N,L,m)=(0,8,3,34)$.
}
\label {imaginary}
\end{figure}

The case   with $L=5$, $m=0$,
$N=12$ is depicted in figure~\ref{3string}.
With increasing $m$, the  roots form longer strings and
the number of roots exceeds $(L-2)N$.
Finally all but 4N  zeroes are on the imaginary axis.
The remaining 4N zeroes lie in 4 complex groups, which are
empirically  located near 
$\pm \epsilon \pm i$ for small real part $\epsilon$.
The BAE roots for  $(M,N,L,m)=(0,8,3,34)$ are
plotted in  figure \ref{imaginary}.%
This behaviour will be discussed in 
appendix \ref{Heun} in the light of an explicit solution.

\begin{figure}[hbtp]
\centering
\includegraphics[width=6cm]{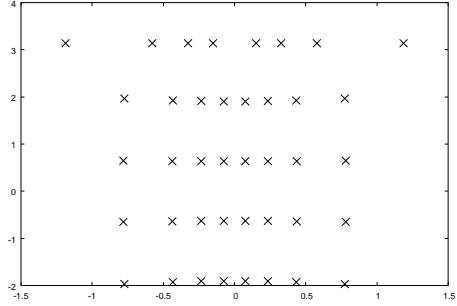}
\caption{ \small
Zeroes of the wavefunction of   (\ref{PT-schro}) with
$(M,N,L,m)=(8,0,5,0)$. We have $M-$ string in the vertical direction.
}
\label {MnezeroNzero}
\end{figure}

Next we consider the case $M \ne 0$ and $N=0$.
When $m=0$, $N_k=0$ and $L$ odd, there are $M$ (almost) strings of
length $L$. The top roots, which are located at $\Im m (u)=\pi$,
are  displaced from the others: the distance between these
roots is a
little bit larger than others.
The example $(M,N,L,m)=(8,0,5,0)$ is shown
in figure~\ref{MnezeroNzero}.
This configuration can also be interpreted as $L$ strings of 
length $M$, rotated by 90$^{\circ}$.
The distance between roots in a string, however, is much less 
than $\frac{2 \pi}{L}$.
In this  interpretation,
the configuration for the case $m=0$, $N_k=0$ and $L$, $M$ even  is similar.
 For  $m=0$, $N_k=0$, $L$ even and $M$ odd, the pattern is
a little bit different; there are $L-2$ strings of length $M$,
a string of length $M-1$ centred at $x= \pi i$ and  a string of length 
$M+1$ centred at $x=0$.

\begin{figure}[hbtp]
\centering
\includegraphics[width=6cm]{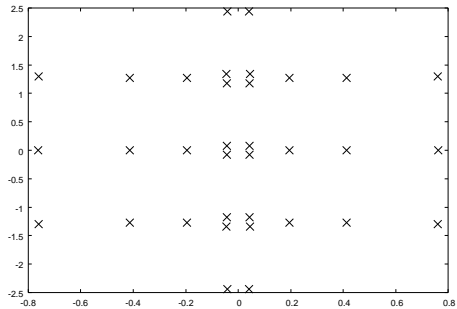}
\caption{ \small The $(M,N,L,m)=(2,8,5,0)$ case.}
\label {s3}
\end{figure}

When both $M$ and $N$ are nonzero, patterns are generally quite involved.
However a very simple picture emerges
for $m=0$: the coexistence of
 the $L-2$ strings, symmetric with respect to the real axis, and
 $M$ strings  symmetric with respect to the imaginary axis.
The example of $L=5$, $M=2$, $N=8$, $m=0$ is shown
in figure~\ref{s3}.

\resection{The field theory limit}
\label{cl}
It is interesting to find   the field theory limit of our
systems.  We shall  work directly with
equation (\ref{odegauge}), and   send $ N \rightarrow \infty$  keeping
$L$, $M$ and $m$ finite. At the same time we focus on
the region near $x=0$ by introducing a new variable  $z$ via
\eq
x^L= \frac{z^L}{L^2 N(N+1)}
\en
and confine our analysis to the region
$|z^L| \ll   L^2 N(N+1)$.
Taking
the large $N$ limit, $\psi_{M,N,m}(z) \rightarrow  \psi_{m}(z)$  with
\eq
\frac{d^2 \psi_m(z)}{dz^2} - \frac{m}{z} \frac{d\psi_m(z)}{dz} -
 z^{L-2} \psi_m(z)=0~.
\label{lairy}
\en
Notice that at  $m=0$ and $L=3$,  equation (\ref{lairy}) coincides
with the Airy equation~(\ref{airy}). However, in order to identify
the continuous  limit  of the Q function at  $m=0$, $ L=3$
with  the Airy function and hence with result of \cite{DT1}
from the ODE/IM  correspondence,
we   should also check its  asymptotic behaviour.
(A possible  difference in overall normalisation will be ignored.)
This is the point  were the boundary condition (\ref{zp}) becomes important.
First notice that in terms of  $z$
the point  where the condition
\eq
\psi_{M,N,m}(z)|_{z \sim z_0} \sim (z_0-z)^{N+1}
\label{bcz}
\en
was imposed (see equation (\ref{zp}))
is   $z_0=(L^2 N(N+1))^{1/L}$, so as $N$ is increased $z_0$  moves toward
infinity.
At first one  might be tempted to extract the large $z$ asymptotic
behaviour of  $\psi_{M,N,m}(z)$ by studying the large $N$ limit of (\ref{bcz}).
Since we already restricted ourselves to
the region $|z^L| \ll   L^2 N(N+1)$  this would be  incorrect:
near the point $z=z_0$
($x=1$)  a linear approximation for the ``potential''
\eq
P(x)=
\left(\frac{L^2  N(N+1) x^{L-2}}{(x^L-1)^2}
  -\frac{L^2  M(M+1) x^{L-2}}{(x^L+1)^2}
   \right )
\en
is clearly unreasonable.
However it is straightforwardly proved that, for $m=0$ and $N \ge M$, the
condition (\ref{bcz}) constrains  $\psi_{N,M,m}(z)$ to be monotonically
decreasing in   the   whole segment
\eq
0 <  z \le  (L^2 N(N+1))^{1/L}~.
\en
This property  guarantees that the  purely subdominant solution is
singled out from  (\ref{lairy}),
giving $q(z) \rightarrow \psi_0(z) \propto  \hbox{Ai}(z)$.
The argument is the following.

The condition (\ref{bcz}) means that
\eq
\psi_{M,N,m}(x)|_{x = 1-\varepsilon} > 0~,
~~~\psi_{M,N,m}'(x)|_{x = 1-\varepsilon} < 0~,
~~~\psi_{M,N,m}''(x)|_{x = 1-\varepsilon} > 0~,
\label{bcz1}
\en
with a small but finite positive  $\varepsilon$.
So decreasing $x$ slightly  below $1$,  $\psi(x)$ remains positive and
in order to change the sign of $\psi'(x)$  the sign of  $\psi''(x)$
 should become negative  first.
Note now that for  $m=0$
\eq
\psi_{M,N,m}''(x) =P(x) \psi_{M,N,m}(x)~,
\label{sec}
\en
and that $P(x)$ is positive  in  $0<x<1$ for $N \ge M$. Then the only way to
have $\psi''(x)=0$ is through  $\psi(x)=0$.
By continuity from $x=1$, this contradicts the positivity
condition~(\ref{bcz1}).

For $m>0$, due to the presence
of the first derivative term in (\ref{odegauge}), this simple argument
does not immediately apply. However, by slightly more involved reasonings
 one can argue that at least for moderate values of  $m$ and  $M$
it is
always the
subdominant solution which  is  singled out in this field theory  limit.
For example the  $L=3$ and $m=1$ case related to (\ref{FSZ}) of \cite{FSZ}
leads to $q(x) \rightarrow \psi_1(z) \propto \hbox{Ai}'(x)$.

Finally, we would like to mention that
for  $m=0$ and   $L$ general the limiting equation (\ref{lairy})
coincides, up to a trivial change of variable, with  the
$\alpha=1$, $l=0$,  $S=(L-2)/2$  case  of the   equation
\eq
-\frac{d^2}{dx^2} \chi(x,E) + \left((x^{\alpha}-E)^{2S} +\frac{l(l+1)}{x^2}
\right)  \chi(x,E)=0~,
\label{gsch}
\en
which has been identified by Lukyanov~\cite{Lsem} with the scaling limit
of the spin$-\frac{L-2}{2}$ XXZ quantum chain.

\resection{Summary and discussion}
\label{con}
In this paper, the generalised  P{\"o}schl-Teller (Heun) equation
in the complex plane has been addressed in view of the connection
relation.
Remarkably, at particular values of parameters, a hidden link to
one dimensional quantum  systems of higher spins has been found.
The Bethe ansatz method, well-developed in the theory of
quantum integrable systems, then provides a simple characterisation
of the wavefunction as an entire function.
These results  begin to fill a gap in earlier studies:
they show that the ODE/IM correspondence has a r\^ole to play in at
least some finite lattice systems.
The place of massive theories in this story, however,
is yet to be clarified.

We would like to remark that  a
very recent investigation of the deformed nonlinear $\sigma$
model\cite{FateevOnofri} establishes a connection between
perturbed $Z_N$ parafermion theory (a massive theory)
and the Heun equation. Though the r{\^o}le played  by the ODE in the
context\cite{FateevOnofri}  is quite far  from the spirit of the  ODE/IM
correspondence, it would be nice to see
whether the analysis proposed  in
this paper could  tell us anything
interesting  about the problem~\cite{FateevOnofri}.

Finally, we add some further comments on the implications of our 
results for the quantum spin chain problem.
The connection rule for the ODE makes full use of the peculiarity of
$q=e^{i \lambda}$ being a root of unity, which can be naturally
extended from $L=3$ to integer values of $L$.
Correspondingly, some peculiar features of spin model with  $S=\frac{1}{2}$
are inherited by spin models for which the quantum space possesses
higher spins, while the spin of the auxiliary space remains at $\frac{1}{2}$.
Physically, vertex models,
or the corresponding Hamiltonians, for which the quantum and  the
auxiliary spaces
share the same magnitude of spin are more relevant. We call these
``isotropic''.
Then a  natural question arises: can we observe  a similarly-simple
behaviour in  the largest eigenvalues of isotropic
transfer matrices of higher spin chains,
  $T\sim ({\rm const.})^{{\cal N}_S}$, with a proper choice of
coupling constant?
Our preliminary numerical results answer this positively.
Under periodic boundary conditions, the 19 vertex model ($L=4$)
and the 44 vertex model ($L=5$) show the desired simple behaviour
for ${\cal N}_S=3,5,7$ when  $\triangle = -\cos\frac{\pi}{L}$.
Indeed, for $L=5$, this is confirmed by the result in section \ref{model}
and fusion relations.
Although these eigenvalues are characterised by the same BAE solutions
as in 
the anisotropic cases, for the isotropic models they turn out to be 
special members of the spectrum -- in fact, the   largest in the
given spin sector.
We also investigated Hamiltonians with  free-free boundaries.
Through numerical diagonalisation, the spin 1 chain  ($L=4$) with
quantum group invariant boundaries\cite{PS} turns out to
possess the ground state energy $E_0 = -4({\cal N}_S-1) $ for
$\triangle = \cos\frac{\pi}{4}$. This is exactly  the expected
behaviour if $T$ obeys the power law.
The origin of this peculiarity, associated to the spin chain,
has been argued for $S=\frac{1}{2}$
to be the representation theory of the quantum algebra
\cite{Hinrichsen,Martin-Sierra}.
There is also an interesting  relationship  between the
antiferromagnetic spin-$\frac{1}{2}$ XXZ quantum  chain  at
$\Delta=-1/2$ and a supersymmetric model of hard-core
fermions~\cite{FendleySchoutensNienhuis}. It is conceivable that
most of the special properties emerging from our analysis
will ultimately find a  natural interpretation  in the framework
of similar supersymmetric systems.

We conclude this discussion by  noting that the ODE/IM
correspondence has been extended in~\cite{DT3,JS4,DDT2,BHK} to
higher order differential equations. In these papers   a
relationship  between $n^{\rm th}$-order ODEs and the conformal limit of
$SU(n)$ lattice  models was established. It is interesting that
these more complicated families of systems
also possess
exactly solvable points. (The corresponding differential equations
are direct generalisations of the Airy equation~(\ref{airy}).)
At least for these cases, the finite lattice extension of the
models should  be straightforward, and we hope to explore this point
further in a future publication.

\bigskip

\noindent
{\bf Acknowledgements}
\medskip

\noindent
We are grateful to V. Bazhanov, G. Falqui, S. Lukyanov  and W.-L.
Yang  for useful discussions. PED and RT thank  Shizuoka
University and YITP-Kyoto, and JS thanks the Mathematics
Department of Durham University,  for hospitality  while some of
this work was in progress. RT thanks the EPSRC for an Advanced
Fellowship. This work was partially supported by the EC network
``EUCLID", contract number HPRN-CT-2002-00325. The
work of PED was also supported in part by JSPS/Royal Society grant
and by the Daiwa Foundation, while
the work of JS was supported by a Grant-in-Aid
for Scientific Research from the Ministry of Education, Culture,
Sports and Technology of Japan, no.~14540376.

\appendix
\section{A solution in terms of the Heun function }
\label{Heun}
The ODE (\ref{PT-schro}) has an explicit solution in terms of
the Heun series $u=H(d,e;\alpha,\beta, \gamma,\delta; z)$ which satisfies

\begin{equation}
 \frac{d^2 u}{dz^2} +
  \bigl( \frac{\gamma}{z}+ \frac{\delta}{z-1}+\frac{\epsilon}{z-d} \bigr ) 
  \frac{d u}{dz}+
  \frac{\alpha \beta(z-e)}{z(z-1)(z-d)} u =0~,
\label{heun}
\end{equation}
with $\alpha, \beta, \gamma,\delta, 
\epsilon:=\alpha+\beta-\gamma-\delta+1, d, e \in \mathbf{C}$.
(Note that $e$ is denoted by $q$ in \cite{whittakerwatson}.)
There are four regular singularities at $z=(0,1,d,\infty)$.
It is well known that any Fuchsian function with  four regular singular points
can be transformed
into the Heun function.

Let us make the
connection between
$\widehat{\psi}$ in  (\ref{fundametals}) and the Heun function.
Set
\eq
x^{- \sigma L/2} \widehat{\psi}_{M,N,m}(x)= \frac{\xi}{\tanh^{M} s 
~\cosh^{\sigma} s}
~,~~~
 (\sigma= {(m+1)/ L})~,
\en
and adopt a variable
$z=x^L=-\exp(2 s)$.
It is then easily established
that
\eq
\xi=(z-1)^{\alpha}  H(d,e;\alpha,\beta, \gamma,\delta; z)~,
\en
with parameters
\bea
\alpha=\sigma-M-N~,~~~
&\beta=-M-N~,&~~~
\gamma=1+\sigma~,~~~\delta=-2N~, \nn \\
\epsilon=-2M~,~~~
&d=-1~,&~~~e=\frac{M-N}{N+M}~.
\label{paraheun}
\eea

The case $M=0$ and $m \gg 1 $ was numerically investigated
in the main text.
In this limit, it is immediate to see,
by its degeneration to the hypergeometric function,
that $H_{M,N,m}(x^L) \rightarrow  (1-x^L)^N$.
 Thus
\eq
\psi_{M,N,m}(x) \rightarrow x^{(m+1)/2} ( x^{(m+1)/L} -  x^{-(m+1)/L} (-1)^N)~,
\en
which explicitly supports the
asymptotic locations of the zeroes being on the unit circle, or
on the imaginary axis  in terms of  $u$.
However this does not account for the fact that most of them are exactly on
the imaginary axis for large but finite $m$.

\resection{The determination of locations of trivial zeroes of 
the wavefunction}
\label{fixtrivialzeros}

In this appendix, we
explain how to locate the `trivial' zeroes of the
wavefunction.
Taking into account the singularities  (\ref{poles}),  there must
be special    solutions of  the form
\eq
\widehat{\psi}_{M, N, m}(x^L) = \frac{x^{ \frac{(2 \sigma-M-N)L}{2}} }
     {(x^{L/2}+x^{-L/2})^M   (x^{L/2}-x^{-L/2})^N    }
H_{M,N,m}(x^L)\,,
\label{fundametals}
\en
where $H_{M,N,m}(z)$ is nonsingular and, as a consequence of
(\ref{rotate2}), should satisfy
\eq
 H_{M,N,m}(-z) \propto  H_{N, M ,m}(z)~.
\label{hprop}
\en

As discussed in appendix \ref{Heun}, $H_{N, M ,m}(z)$ can be
written  in terms of the Heun function. In view of this explicit
form, property  (\ref{hprop}) is immediate: note first that
$H_{M,N,m}(x^L)= H(-1,e,\alpha,\beta,\delta;z)$. If $d=-1$,
(\ref{heun}) is invariant under $z \rightarrow -z$, $e\rightarrow -e$,
 $\epsilon \leftrightarrow \delta$.
This is  accomplished in our case by $x^L \rightarrow -x^L$,
$M\leftrightarrow  N$, which can be easily verified using the
 parameterisation (\ref{paraheun}) in terms of $\sigma$,  $M$ and $N$.
Thus the desired property   (\ref{hprop}) is shown to be valid.

Using  the symmetry    $s \rightarrow -s$  of the original
P{\"o}schl-Teller  equation (\ref{PT-schro})
one can check    that
\eq
x^{\sigma L} \widehat{\psi}_{M, N, m}(1/x^L)
\en
is also a   solution of (\ref{odegauge}),
which turns out to be independent of
(\ref{fundametals}) so long as  $\sigma \notin \mathbf{Z}$.
Therefore the particular solution which satisfies  the boundary condition
(\ref{zp}) is
\bea
\psi_{M, N, m}(x)&=& \widehat{\psi}_{M, N, m}(x^L) -
 (-1)^N  x^{\sigma L}  \widehat{\psi}_{M, N, m}(\frac{1}{x^L})
\nn\\[3pt]
& =  &
\frac{1} {(x^{L/2}{+}x^{-L/2})^M   (x^{L/2}{-}x^{-L/2})^N    } \nn
\\[3pt]
&&{}\times
      \Bigl(   x^{ \frac{(2 \sigma-M-N)L}{2}}  H_{M,N,m}(x^L)
             -   x^{ \frac{(M+N)L}{2}}  H_{M,N,m}(\frac{1}{x^L})\Bigr)
~.
\label{psisol}
\eea
{}From the expression (\ref{psisol}) one can
read off the   positions and orders
of  the trivial zeroes and poles.
Consider first zeroes related  to the symmetry  (\ref{rotate1}).
We denote the greatest common divisor of $m+1$ and $L$ by
\eq
\ell={\rm GCD} (m+1, L)~,
\label{ell}
\en
 and set
\eq
\omega =\exp(2i\eta)~,~~ \eta=\frac{\pi}{L}~~~~\hbox{and}~~~
\omega'= \exp(2i \eta')~,~~  \eta'=\frac{\pi}{\ell}~.
\label{omega1}
\en
Then at $x=\omega'^{k},
(k=0,1,\cdots, \ell-1)$,  $\psi_{M,N,m}(x)$ has zeroes of order $N+1$,
and  at $x=\omega^{k}, (k=1,\cdots, L-1)$ 
such that ${\rm GCD} (k\ell, L)=1$,
it has poles of order $N$.

There are also   zeroes and poles
related  to  the symmetries  (\ref{rotate2}).
Let $k_i$ be a positive integer satisfying
\eq
( 2 k_i+1) \bigl (  N+M -\frac{m+1}{L})  \in  2 \mathbf{Z}~,~~~
 (1\le k_i \le L-1)~.
\label{ki}
\en
By paying attention to  (\ref{psisol}), especially  the balance of 
the two terms
in the numerator,
we  check that  $x=\omega^{(2k+1)/2}$ is a pole of
the order $M$ of $\psi_{M,N,m}(x)$  if $k \ne k_i$, while it is a
zero  of the order $M+1$ when  $k = k_i$.
One can easily check that if $k_i$ is a solution of
(\ref{ki}) then so also is $k_i+\frac{L}{\ell}$.
The number of
possible $k_i$, $N_k$, is  thus either zero or $\ell$.
%
%

%
\end{document}